\documentclass[5p]{elsarticle}
\usepackage{xcolor}
\usepackage{lineno,hyperref}
\usepackage{amsmath}\modulolinenumbers[5]
\journal{Physics Letters B}

\begin{document}
\begin{frontmatter}

\title{$\Sigma^0$ production in proton nucleus collisions near threshold}

\author[4]{J.~Adamczewski-Musch}
\author[7]{G.~Agakishiev}
\author[9,10]{O.~Arnold}
\author[15]{E.T.~Atomssa} 
\author[8]{C.~Behnke}
\author[9,10]{J.C.~Berger-Chen} 
\author[3]{J.~Biernat}
\author[2]{A.~Blanco}
\author[8]{C.~Blume}
\author[10]{M.~B\"{o}hmer}
% ...

\address{(HADES Collaboration)}
% ...
\address[2]{LIP-Laborat\'{o}rio de Instrumenta\c{c}\~{a}o e F\'{\i}sica Experimental de Part\'{\i}culas , 3004-516~Coimbra, Portugal}
\address[3]{Smoluchowski Institute of Physics, Jagiellonian Universityof Cracow, 30-059~Krak\'{o}w, Poland}
\address[4]{GSI Helmholtzzentrum f\"{u}r Schwerionenforschung GmbH,64291~Darmstadt, Germany}
% ...
\address[7]{Joint Institute of Nuclear Research, 141980~Dubna, Russia}
\address[8]{Institut f\"{u}r Kernphysik, Goethe-Universit\"{a}t, 60438~Frankfurt, Germany}
\address[9]{Excellence Cluster 'Origin and Structure of the Universe' ,85748~Garching, Germany}
\address[10]{Physik Department E62, Technische Universit\"{a}t M\"{u}nchen, 85748~Garching, Germany}
% ...
\address[15]{Institut de Physique Nucl\'{e}aire (UMR 8608), CNRS/IN2P3 -Universit\'{e} Paris Sud, F-91406~Orsay Cedex, France}
% ...
\address[c]{ also at Technische Universit\"{a}t Dresden, 01062~Dresden,Germany}
% ...

\begin{abstract}
The production of $\Sigma^{0}$ baryons in the nuclear reaction p (3.5 GeV) + Nb (corresponding to $\sqrt{s_{NN}}=3.18$ GeV) is studied with the detector set-up HADES at GSI, Darmstadt. 
$\Sigma^{0}$s were identified via the decay $\Sigma^{0} \rightarrow \Lambda \gamma$ with subsequent decays $\Lambda \rightarrow p \pi^{-}$ in coincidence with a $e^{+}e^{-}$ pair from 
either external ($\gamma \rightarrow e^{+} e^{-}$) or internal (Dalitz decay $\gamma^{*}\rightarrow e^{+} e^{-}$) gamma conversions.
The differential $\Sigma^0$ cross section integrated over the detector acceptance, i.e. the rapidity interval $0.5 < y < 1.1$, has been extracted as 
$\Delta\sigma_{\Sigma^{0}} = 2.3 \pm (0.2)^{stat} \pm \left(^{+0.6}_{-0.6}\right)^{sys} \pm (0.2)^{norm}$ mb, yielding the inclusive production cross section in full phase space 
$\sigma^{total}_{\Sigma^{0}} = 5.8 \pm (0.5)^{stat} \pm \left(^{+1.4}_{-1.4}\right)^{sys} \pm (0.6)^{norm} \pm (1.7)^{extrapol}$ mb by averaging over different extrapolation methods.
The $\Lambda_{all}$/$\Sigma^{0}$ ratio within the HADES acceptance is equal to 2.3 $\pm$ $(0.2)^{stat}$ $\pm$ $(^{+0.6}_{-0.6})^{sys}$. The obtained 
rapidity and momentum distributions are compared to transport model calculations. The $\Sigma^{0}$ yield agrees with the statistical model of particle production in nuclear reactions.
\end{abstract}

\begin{keyword}
Hyperons, Strangeness, Proton, Nucleus
\end{keyword}
\end{frontmatter}

%\linenumbers
\emergencystretch 1cm
\section{Introduction}
The study of hyperon production in proton-induced collisions at beam energies of a few GeV is important for many open questions in the field of hadron physics. While several 
experimental results exist for $\Lambda$ hyperons in p+p and p+A reactions \cite{Adamczewski-Musch:2016vrc,Munzer:2017hbl,Agakishiev:2014kdy,Kowina:2004kr,Sewerin:1998ky,Sullivan:1987us}, 
measurements of $\Sigma^0$ production are scarce \cite{Kowina:2004kr,Sewerin:1998ky,Sullivan:1987us}. The dominant electromagnetic decay $\Sigma^{0}\rightarrow \Lambda +\gamma$ (BR $\approx$ 100$\%$) 
requires the identification of photons with E$_\gamma$ $\simeq 80$ MeV concident to the detection of p$\pi^-$ pairs from $\Lambda$ decays. 
Our measurement is the first step towards gainig access to the hyperon electromagnetic form factors \cite{Leupold}. Once the measurement of virtual photons in the Dalitz decay $\Sigma^0 \rightarrow \Lambda e^+e^-$  (BR $<$ 1$\%$)
is performed it can be separated from the decays involving a real photon and therefore provide complementary information on the nucleon and $\Delta$ baryon form factors \cite{Adamczewski-Musch:2017hmp}.\\
Hadron collisions at energies of a few GeV with hyperons in the final state are also suited to study the role played by intermediate hadronic resonances in the strangeness production process. 
Indeed, non-strange resonances like N* and $\Delta$ have been found to contribute significantly \cite{AbdelSamad:2006qu,Roder:2013gok,Siebert:1994jy,Hauenstein:2016zys,Agakishiev:2011qw} 
via the channels $N^*\rightarrow \Lambda +K^+$ and $\Delta^{++} \rightarrow \Sigma(1385)^+ +K^+$. In case of N*, up to seven resonances with similar masses and widths have been identified including 
the occurrence of interference effects among them \cite{Munzer:2017hbl,Agakishiev:2014dha}. In this context, the simultaneous measurement of $\Lambda$  and $\Sigma$ hyperons becomes important 
to understand the interplay between the spin $1/2$ and $3/2$ states occurring in the strong conversion process $\Sigma+N \rightarrow \Lambda +N$. This process manifests itself as a 
peak structure on top of the smooth $\Lambda + p$ invariant-mass distribution close to the $\Sigma$-$N$ threshold and is known to be responsible for cusp effects \cite{AbdEl-Samad:2013ida}.
Hyperon production in nuclear reactions gives also access to details of the hyperon-nucleon interaction. The existence of $\Lambda$ hypernuclei is argued as evidence for an attractive potential 
at rather large inter-baryon distances \cite{Feliciello:2015dua,Gal:2016boi}. Theoretical models \cite{Haidenbauer:2013oca} trying to describe scattering data \cite{SechiZorn:1969hk,Eisele:1971mk}
with hyperon beams postulate the presence of a repulsive core for the $\Lambda$-$N$ interaction. $\Sigma^{0}$ hypernuclei, on the other hand, have not been observed so far due to difficulties implied 
by the electromagnetic $\Sigma^{0}$ decays and the requirement of large acceptance and high resolution electromagnetic calorimeters. Since also scattering data for $\Sigma$ hyperon beams are scarce, constraints
on the $\Sigma$-N interaction are missing so far and new measurements of $\Sigma^{0}$ production in nuclear targets are essential.\\
Medium-energy heavy-ion collisions producing hyperons allow to study their properties within a dense baryonic environment (up to $\rho\approx 2-3\rho_0$) 
\cite{Bastid:2007jz,Agakishiev:2010rs,Pinkenburg:2001fj,Chung:2001je}. One question of interest is whether the attractive $\Lambda$-N interaction in vacuum or at nuclear saturation might change due
to the postulated appearence of a more dominant repulsive core at increased densities and short distances \cite{Petschauer:2015nea}. The quest for  detailed information on such aspects requires 
the knowledge of $\Lambda$ feed down effects from $\Sigma^0$ production and its corresponding behaviour in baryonic or even cold nuclear matter.\\
Experimental data for simultaneous $\Sigma^{0}$ and $\Lambda$ production are available 
for proton-proton collisions either close to the free NN production threshold (E$_{th}$ = 2.518 GeV for $\Lambda$ and E$_{th}$ = 2.623 GeV for $\Sigma^0$) \cite{Kowina:2004kr,Sewerin:1998ky} or at  
excess energies of $\simeq 5\,$GeV and above \cite{LandoldBoernstein}. So far, no data are available for $\Sigma^{0}$ hyperons emerging from proton + nucleus collision systems at few GeV incident 
beam energy. In this work we present the first measurement of $\Sigma^0$ production in p + Nb collisions at an incident kinetic beam energy of $E_{p}$ = 3.5 GeV.  Our paper is organised as follows. In section 2, we describe the experimental set-up. Section 3 is devoted to $\Sigma^0$ identification and background subtraction. In section 4 the method for efficiency correction and differential analysis is shown. In section 6 the extracted cross sections and yields are compared to different models. In sections 6 we give a summary and short outlook.

\section{The HADES experiment}
The High-Acceptance Di-Electron Spectrometer (HADES) \cite{Agakishiev:2009am} located at the GSI Helmholtzzentrum f\"ur Schwerionenforschung in Darmstadt (Germany) is an experimental facility for 
fixed target nuclear reaction studies in the few GeV energy region. The spectrometer is dedicated to measure low-mass dielectrons originating from the decay of vector mesons in the invariant-mass 
range up to the $\phi$ mass and offers excellent identification by means of charged hadrons such as pions, kaons and protons. The detector setup covers polar angles between 18$^{\circ}$ to 85$^{\circ}$ over almost 
the full azimuthal range designed to match the mid-rapidity region of symmetric heavy ion collisions at E = 1-2 AGeV. A set of multi-wire drift chamber (MDC) planes arranged in a sixfold segmented 
trapezoidal type structure, two layers in front and two behind a toroidal magnetic field, is used for charged-particle tracking and momentum reconstruction with a typical resolution of $\Delta$p/p $\simeq$ 3$\%$. 
An electromagnetic shower detector (Pre-Shower) and a Time-Of-Flight scintillator wall (TOF and TOFINO) build the Multiplicity and Electron Trigger Array (META) detector system used for event trigger 
purposes. The energy loss (dE/dx) signals measured in the TOF and MDC detectors are used for charged particle identification. In addition, electrons and positrons are identified over a large range of 
momenta with a Ring Imaging Cherenkov (RICH) detector surrounding the target in a nearly field-free region.\\
In the present experiment, a proton beam accelerated by the SIS18 synchrotron to a kinetic energy of E$_p$ = 3.5 GeV has been directed on a twelve-fold segmented  $^{93}$Nb target of 2.8\% nuclear interaction 
probability. For a TOF+TOFINO reaction trigger setting of multiplicity M $\geq$ 3 and at typical beam intensities of $2\times 10^{6}$ particles/s on target, a total of 3.2 $\times$ 10$^{9}$ events have been 
recorded and analyzed.
 
\section{$\Sigma^{0}$ identification and background subtraction}
\begin{figure}  
   \includegraphics[width=0.45\textwidth]{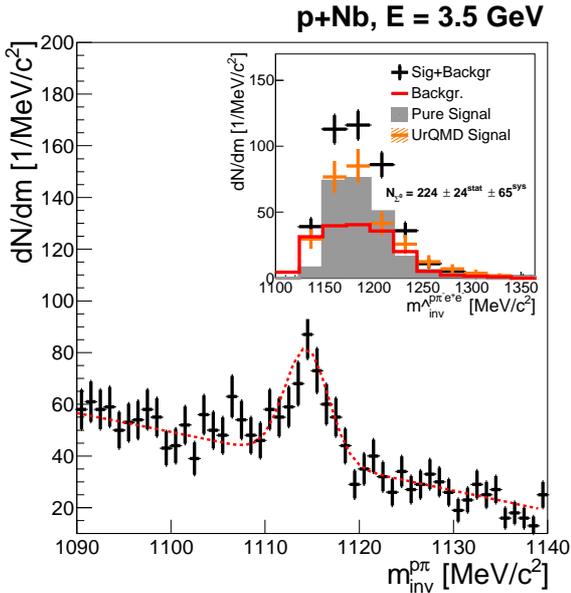}
   \caption{(Color online) Invariant mass distribution of p$\pi^-$ pairs with an additional e$^{+}$-e$^{-}$ pair in the same event. The dashed curve shows a combination of a polynomial background fit and a 
   gaussian fit applied to the signal area. {\bf Inset:}  Four-particle invariant mass distribution of a proton, pion and dielectron for p$\pi^{-}$ pairs in the  $\Lambda$ signal region. Measured signal 
	    (black crosses), combinatorial background (red histogram) and extracted net signal (gray line) are shown in comparison to a UrQMD simulation (orange histogram) with scaled $\Sigma^{0}$ production.}
   \label{fig:pip}
\end{figure}

The identification of $\Sigma^{0}$ hyperons was achieved via the decay channel $\Sigma^{0} \rightarrow \Lambda \gamma$ (BR $\approx$ 100 \% \cite{PDG}) by reconstructing $\Lambda \rightarrow p \pi^{-}$ 
decays correlated with the emission of a dielectron from external pair conversion $\gamma \rightarrow e^{+} e^{-}$ or from the Dalitz decay $\Sigma^{0} \rightarrow \Lambda e^{+}e^{-}$ 
(BR $< 5 \cdot 10^{-3}$ \cite{PDG}). Figure \ref{fig:pip} depicts the p$\pi^{-}$ invariant-mass distribution of such events with a clear signature of a $\Lambda$ content in the data sample. 
Due to the low mass difference $m_{\Sigma^{0}}-m_{\Lambda} \approx$ 77 MeV/c$^{2}$ a considerable fraction of coincident e$^{\pm}$ candidates have momenta below the spectrometer acceptance 
threshold p$_{thr} \approx 50$ MeV/c needed for full track and momentum reconstruction. For this reason, dielectrons have been identified by requiring two RICH rings, at least one fully 
reconstructed e$^{\pm}$ track and one neighbouring incomplete tracklet detected in front of the  magnetic field in the first two MDCs. The missing momentum of the incomplete tracklet has been estimated by applying a 
most probable hypothesis as described in detail in \cite{Kunz:2017} which partly exploits results and constraints from kinematically similar $\pi^0$ Dalitz decays. In this way, the observed incomplete dielectrons 
 are combined into most probable photon signals with a resolution of $\delta$E$_{\gamma}($FWHM$) = 57 \pm 2$ MeV \cite{Kunz:2017}.

The combinatorial background has been determined with two approaches. First, background yield and shape have been estimated 
from polynominal fits of the p$\pi^{-}$  invariant mass  in the sideband regions below and above the $\Lambda$ peak,  
1090 MeV/c$^{2} <$ m$^{p\pi^{-}}_{\textrm{inv}} <$ 1105 MeV/$^{2}$ and 1125 MeV/c$^{2} <$ m$_{\textrm{inv}}^{p\pi^{-}} <$ 1140 MeV/c$^{2}$ 
(see Fig.\ref{fig:pip}). 
The second approach aimed at the suppression of a random peak structure. The momentum of the proton and pion was smeared by 2\% such that the resulting invariant mass of proton and pion did not show
any $\Lambda$ peak. The obtained distribution was scaled to the sideband of the unsmeared distribution shown in Fig.\ref{fig:pip} to evaluate the background in the signal region.
After weighting and normalisation, both methods lead to the same background yield within 10\%. 
The side band samples used to construct the p$\pi^-$ background are combined with the reconstructed e$^{\pm}$ pairs to obtain
the background to the $\Sigma^0$ candidates (for details see \cite{Kunz:2017}).\\
The inclusive four-particle p$\pi^-$e$^{\pm}$ invariant mass distribution is shown in the inset
 of Fig.\ref{fig:pip}.  
A peak structure becomes apparent at the $\Sigma^0$ pole mass with a width (FWHM) of $52 \pm 22$ MeV/c$^{2}$. 
The observed FWHM is mainly attributed to the resolution of the $\gamma$ reconstruction.  The estimated background
is shown by the red histogram. \\
Full scale UrQMD~\cite{UrQMD,UrQMD_Version} simulations have been carried out and processed through Geant and a digitisation
procedure to emulate the detector response. Subsequently the events 
 have then been analyzed in the same manner as the experimental data. Then the simulation has been normalised to the $\Sigma^0$ yield. The inset in Fig.\ref{fig:pip} 
shows that the simulated $\Sigma^0$ mass distribution is in agreement with the measured distribution.
 A total of $N_{\Sigma^{0}} = 224 \pm 24^{stat} \pm 65^{sys}$ $\Sigma^{0}$ candidates has been extracted.\\ 
\section{Efficiency correction and differential analysis}
After background subtraction, a differential analysis has been performed for the kinematic variables transverse momentum p$_{t}$ and rapidity $y$ of the $\Sigma^{0}$ candidates. Due to the limited event statistics, the experimental 
yields are computed for three equally spaced momentum bins between $240 MeV/c \le $ p$_{t} \le 960$\,MeV/c split in two rapidity bins $0.5<y<0.8$ and $0.8<y<1.1$. The acceptance and efficiency correction 
matrix for this phase space region has been obtained from simulations utilizing the UrQMD/Geant3 data set (see above) before and after $\Sigma^{0}$ reconstruction. The systematic errors of these corrections
stem from various sources. The uncertainty on particle identification of protons and pions of $\simeq 5\%$ is adopted from the high statistics analysis of inclusive $\Lambda$ production \cite{Agakishiev:2014kdy}. 
The overall uncertainty for identification of low momentum e$^{+}$/e$^{-}$ partners and pair reconstruction with two complete tracks is $\simeq 25\%$ as deduced in a previous search for 
dark photons with hypothetical masses in the interval $\simeq$ 50-100 MeV \cite{DarkPhoton}. The error in the background subtraction is estimated from a comparison of the two methods described above and 
contributes with $\simeq 8\%$. 
Other sources are of order $10^{-2}$ and less. The quadratic sum results in a total systematic error of $\approx$ 30\%. The statistical errors did 
not exceed values of $\simeq 10-30\%$. 

\begin{figure}  
   \includegraphics[width=0.45\textwidth]{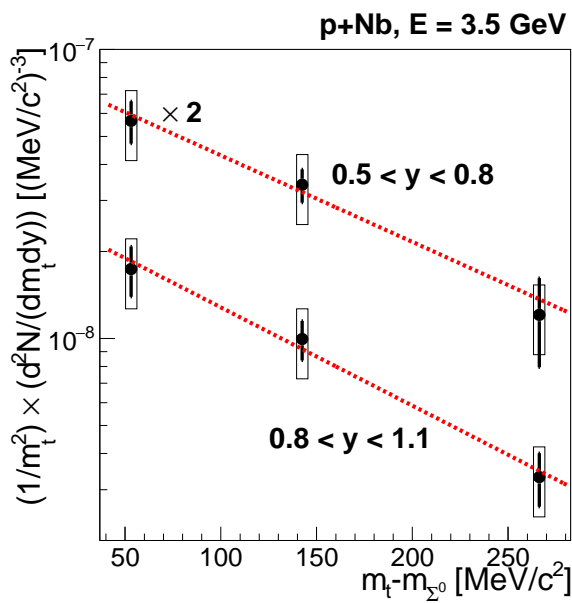}
   \caption{(Color online) Reduced transverse mass distributions of $\Sigma^0$s corrected for acceptance and detection efficiency. The data are plotted for two rapidity bins. The red dashed lines indicate Maxwell-Boltzmann 
	    fits. See text for details.}
   \label{fig:sigma}
\end{figure}

The corrected reduced transverse-mass spectra (with $m_t= \sqrt{p_t^2+m_{\Sigma^{0}}^2}$) for the $\Sigma^0$ candidates are shown in Fig. \ref{fig:sigma} separately 
for both rapidity intervals.
\begin{figure}
   \includegraphics[width=0.5\textwidth]{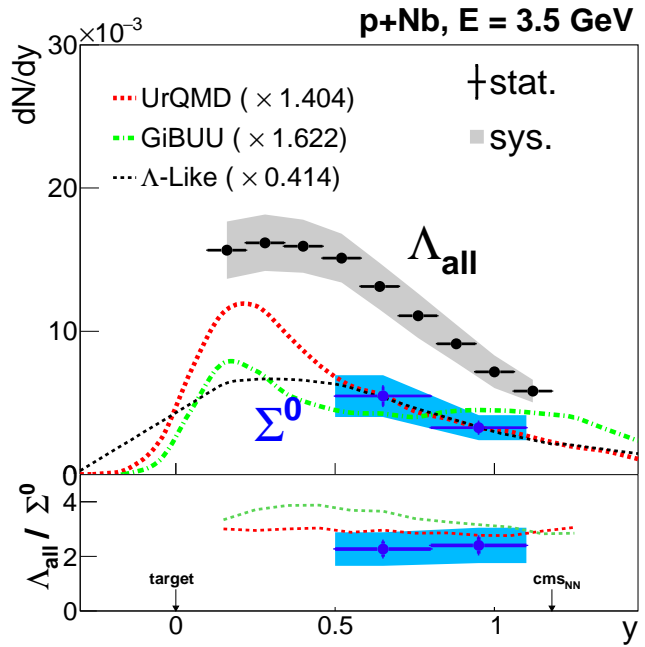}
   \caption{(Color online) {\bf Top:} Experimental rapidity-density distributions of $\Lambda$ (black) and $\Sigma^{0}$ (blue) hyperons. The $\Lambda$ distribution~\cite{Agakishiev:2014kdy} refers to all experimentally identified $\Lambda$s. The shaded bands denote the systematic errors. The dotted lines represent model calculations scaled to 
		       match the measured $\Sigma^0$ yield (see text). {\bf Bottom:} The unscaled ratio $\Lambda_{all}/\Sigma^{0}$. Color and line codes as in top panel.}      
    \label{fig:ratio}
\end{figure}
Towards smaller transverse momenta, the geometrical spectrometer acceptance does not cover the full region for at least one of the decay partners $\Lambda$ or $\gamma$. To extrapolate to uncovered phase 
space regions we have assumed a thermal $\Sigma^{0}$ phase space production. Hence, the differential distributions have been fitted  with a Maxwell-Boltzmann distribution
$(1/m^{2}_{t})(d^{2}N/(dm_{t}dy)) = A(y)\cdot\exp(-((m_{t}-m_{\Sigma^{0}})c^{2})/(T_{B}(y))$, where $A(y)$ is a rapidity dependent scaling factor and m$_{\Sigma^{0}}$ = 1192.642 $\pm$ 0.024 MeV/c$^{2}$ \cite{PDG}. 
The inverse-slope parameters $T_{B}$ = 82 $\pm$ 23 MeV for the rapidity bin 0.5 $<$ y $<$ 0.8 and $T_{B}$ = 78 $\pm$ 22 MeV for the more forward region 0.8 $<$ y $<$ 1.1 
can be compared with the average value of 84 MeV extracted for $\Lambda$ hyperons in the same reaction \cite{Agakishiev:2014kdy}. 

The experimental rapidity-density distributions dN/dy 
obtained for both hyperons from integration of the corresponding Maxwell-Boltzmann distributions with the given parameters are depicted in the upper panel of Fig. \ref{fig:ratio}. 
The calculation of minimum-bias multiplicities requires normalisation of the observed yields to the total number of reactions which we obtained by multiplying the number of M3 triggers
(charged particle multiplicity $\geq\,3$) with a correction factor C. The latter has been extracted from a UrQMD simulation of the p+Nb reaction with impact parameters in the range 0-8 fm 
and full Geant3 propagation of the events yielding  C = 1 / R$_{Trigger}^{M3\rightarrow M1}$ with R$_{Trigger}^{M3\rightarrow M1} = 0.58 \pm 0.06$.
Summation over both rapidity bins in Fig \ref{fig:ratio} gives the multiplicity inside the acceptance N$_{\Sigma^0} = ( 2.7 \pm (0.2)^{stat} \pm \left(^{+0.7}_{-0.7}\right)^{sys} \pm (0.2)^{norm} )\times 10^{-3}$/evt. and
N$_{\Lambda_{all}} = ( 6.1 \pm (^{+0.3}_{-0.3})^{sys} \pm (0.8)^{norm} \times 10^{-3}$/evt. Note that the N$_{\Lambda_{all}}$ signal includes the feed 
down from heavier resonances, mainly from $\Sigma^{0}$ decays. The production ratio inside the acceptance 0.5 $<$ y $<$ 1.1 is found to be 
$\Lambda_{all}/\Sigma^{0} = 2.3 \pm (0.2)^{stat} \pm \left(^{+0.6}_{-0.6}\right)^{sys}$. 

\section{Cross sections and comparison to models}
%The calculation of production cross sections requires normalisation to the number of M3 triggers (charged particle multiplicity $\geq\,3$). UrQMD simulations for p + Nb collisions 
%with impact parameters in the range of 0-8 fm and full Geant3 propagation yield a correction factor C = 1 / R$_{Trigger}^{M3\rightarrow M0}$ with R$_{Trigger}^{M3\rightarrow M0} = 0.58 \pm 0.06$. 
The production cross section has then been obtained by multiplying the multiplicity with the total interaction cross section $\sigma_{pNb}$ = 848 $\pm$ 126 mb for the p + Nb 
reaction \cite{Agakishiev:2013noa,Agakishiev:2012vj} and correcting it for the trigger bias. The acceptance integrated cross section $\Delta\sigma_{\Sigma^{0}}$ which can be obtained from the experimental count rates by
multiplication with the luminosity is found to be equal to $\Delta\sigma_{\Sigma^{0}} = 2.3 \pm (0.2)^{stat} \pm \left(^{+0.6}_{-0.6}\right)^{sys} \pm (0.2)^{norm}$ mb within the rapidity
interval $0.5 < y < 1.1$.\\
Extrapolation to the uncovered rapidity region and extraction of an estimate for the total production cross section have been deduced with the help of transport model calculations.
We have extracted $\Sigma^{0}$ rapidity distributions from UrQMD \cite{UrQMD} and GiBUU \cite{GiBUU,GiBUU_Version} event generators and normalised them to match the experimental data points. The distributions are plotted in
Fig. \ref{fig:ratio} and exhibit considerable differences. Those possibly indicate different weights in the models for the implementation of the slowing down of the $\Sigma^{0}$ which are initially produced at the rapidity of the NN center-of-mass system. While the data 
are well reproduced by UrQMD in the region above $y > 0.4$, the extrapolation to target rapidities seems to be ambiguous. Under the assumption that both hyperons experience comparable emission kinematics due to
their very similar masses we can profit from the larger rapidity coverage and smaller bin sizes of the reconstructed $\Lambda$. Hence, as an alternative guidance we have used the measured $\Lambda$ rapidity density
distribution ($\Lambda$-like) as published in \cite{Agakishiev:2014kdy} and normalised it to the $\Sigma^0$ distribution. 
For comparison, the resulting total $\Sigma^0$ yields and extrapolated production cross sections of the scaled distributions are listed in tab. \ref{tab:yields}.\\

\begin{table} 
 \small{
  \begin{tabular}{c|cc}
      Shape			&	$\Sigma^{0}$ yield per event  &	$\sigma_{\Sigma^{0}}^{total}$ [mb]			\\
      \hline 
      $\Lambda$-like		&	5.2 $\times$ 10$^{-3}$							&	$4.4 \pm 0.4^{stat} \pm 1.1^{sys} \pm 0.5^{norm}$	\\
      GiBUU			&	7.3 $\times$ 10$^{-3}$							&	$6.2 \pm 0.5^{stat} \pm 1.5^{sys} \pm 0.6^{norm}$	\\
      UrQMD			&	8.6 $\times$ 10$^{-3}$							&	$7.3 \pm 0.6^{stat} \pm 1.8^{sys} \pm 0.8^{norm}$	\\		 
  \end{tabular}
}
    \caption{ Total $\Sigma^0$ yields and cross sections after extrapolation under three assumptions. }
  \label{tab:yields} 
\end{table}

\begin{figure}
   \includegraphics[width=0.5\textwidth]{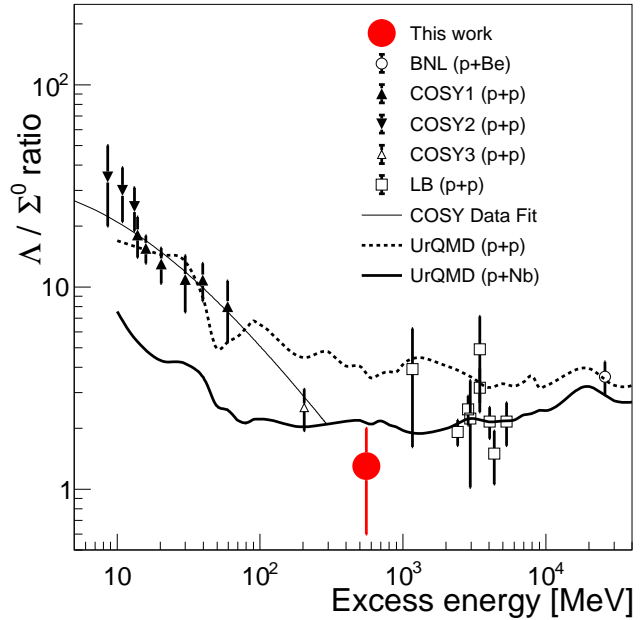}
   \caption{(color online) Experimental excitation function of $\Lambda$/$\Sigma^{0}$ production cross section ratios from exclusive measurements of $\sigma(pp\rightarrow pK\Lambda)$ and 
	    $\sigma(pp\rightarrow pK\Sigma^{0}$) reactions. The excess energy above production threshold refers to free nucleon-nucleon collisions. Data (symbols) from BNL \cite{Sullivan:1987us}, 
	    COSY \cite{Kowina:2004kr,Sewerin:1998ky,AbdelBary:2010pc},
	    LB \cite{LandoldBoernstein} and present work. The thin curve is a fit from \cite{AbdelBary:2010pc}. The dotted and solid curves exhibit UrQMD simulations. Fermi motion has been neglegted for p+A collisions.
	    }
  \label{fig:world}
\end{figure}

The $\Sigma^0$ production cross section has finally been calculated from a mean of the $\Lambda$-like and UrQMD rapidity distributions resulting in
$\sigma^{tot}_{p+Nb}(\Sigma^{0}) = 5.8 \pm (0.5)^{stat} \pm \left(^{+1.4}_{-1.4}\right)^{sys} \pm (0.6)^{norm} \pm (1.7)^{extrapol}$ mb. A $\Sigma^{0}$ yield 
of N$_{\Sigma^{0}}$ = (7 $\pm$ 3) $\times 10^{-3}$/evt for the full phase space has been extracted in the same way. The ratio 
$\Lambda_{all}/\Sigma^{0} = 2.3 \pm (0.2)^{stat} \pm (^{+0.7}_{-0.7})^{sys} \pm (0.7)^{extrapol}$ has been obtained by using the ratio within the acceptance
and an additional extrapolation uncertainty stemming from the difference between UrQMD and $\Lambda$-like extrapolation methods. This can be justified by the rather flat distribution of experimental data as well as for the UrQMD and GiBUU simulations.
The error on the extrapolation procedure introduces the largest uncertainty. The statistical and systematic errors have been added quadratically. 
Figure \ref{fig:world} shows our result for the total number (i.e., full phase space extrapolated) of $\Lambda$s not stemming from $\Sigma^{0}$ decays (that is the number of identified $\Lambda$s minus the number of $\Lambda$s identified as decay products of $\Sigma^0$s) divided by the number of $\Sigma^{0}$s, $R = 1.3\,\pm\,0.6$, together with a compilation of the world data 
\cite{Kowina:2004kr,Sewerin:1998ky,Sullivan:1987us,LandoldBoernstein,AbdelBary:2010pc} and a data fit \cite{AbdelBary:2010pc} plotted as a function of excess energy above the nucleon-nucleon threshold. 
The results from UrQMD are shown for comparison. All data points but two stem from proton-proton collisions. Our result for the production in a heavy nucleus (heavy bullet in Fig. 4) fits well to the 
systematics and model predictions. In this comparison, the multi-step interaction of the $\Sigma^{0}$ with one, two or even more nucleons has been neglected as well as the Fermi motion.\\

We now compare our findings to the statistical model THERMUS \cite{Thermus}. In this model, the total particle 
abundances strictly follow a distribution expected from hadron freeze-out at conditions
determined by a temperature $T_{f.o.}$ and a baryochemical potential $\mu_{f.o.}$. For this 
scenario, particle yields are proportional to $e^{(E-\mu_{f.o.})/T_{f.o.}}$. A THERMUS fit to measured particle yields \cite{HADES_THERMUS}, 
excluding the $\Sigma^0$, gives parameter values 
$T_{f.o.}$ = 100 MeV and $\mu_{f.o.}$ = 620 MeV. 
\begin{figure}
   \includegraphics[width=0.5\textwidth]{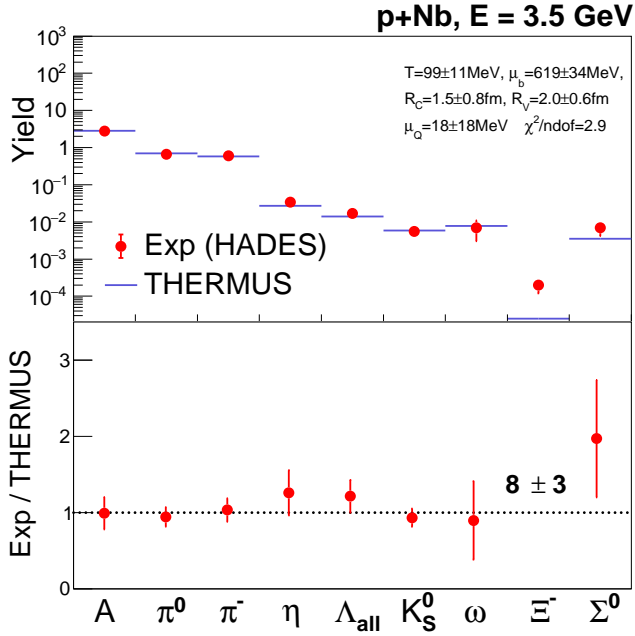}
   \caption{(color online) Experimental hadron yields measured by HADES \cite{HADES_THERMUS} in comparison to a THERMUS statistical model fit w/o $\Sigma^{0}$.}
   \label{thermus}
\end{figure}
For these parameters (see legend in Fig. 5), the expected $\Sigma^0$ yield slightly underestimates 
(1.5 $\sigma$) the inclusive experimental value presented in this work. Figure \ref{thermus} shows the corresponding THERMUS fit results. The THERMUS yield ratio $\Lambda_{all}$/$\Sigma^0 = 3.9$ 
is slightly higher than that predicted by GiBUU, UrQMD (R $\simeq 3$) and our measurement (R $\simeq 2.3$). Nevertheless, the overall agreement 
is surprising for proton induced nuclear collisions 
at relatively low energies, as already discussed in \cite{HADES_THERMUS}.

\section{Summary and Outlook}
We have demonstrated the capability of HADES to reconstruct the low energy $\gamma\rightarrow e^+e^-$ conversion processes
in the detector material via the identification of electrons and positrons.
With this technique we were able to measure for the first time $\Sigma^{0}$ hyperon production in proton-induced reactions off a heavy nucleus near threshold. We provide transverse mass distributions in two rapidity bins.
Based on them, a $\Sigma^{0}$ production cross section of $\sigma_{p+Nb}(\Sigma^{0}) = 5.8~\pm~2.3$ mb has been determined. The inclusive light hyperon production ratio is $\Lambda_{all}$/$\Sigma^{0} = 2.3\,\pm\,1.1$. All uncertainties have been summed up quadratically.
These experimental values compare reasonably well with transport model calculations and results from a statistical hadronisation scheme. In spite of the limited spectrometer 
acceptance the obtained relative production cross sections may hint to a slightly larger production probability in nuclei as compared to expectations from proton-proton collisions, 
$\frac{\Sigma^{0}}{\Lambda}\rvert_{pA} > \frac{\Sigma^{0}}{\Lambda}\rvert_{pp}$. A possible measurement with a low magnetic field will allow full reconstruction of the dielectrons and therefore 
offer the possibility to determine electromagnetic transition formfactors.
The currently ongoing upgrade includes an electromagnetic calorimeter which will significantly enhance the $\gamma$ detection capabilities of HADES. This opens up the investigation
of reaction channels involving photon decays of hyperons and other baryonic resonances produced in proton/pion-proton, proton/pion-nucleus and heavy-ion collisions and might even
give access to measurements of electromagnetic transition form factors for these resonances.

\section*{Acknowledgements}
The HADES collaboration gratefully acknowledges the support by the grants
VH-NG-823, TU Darmstadt (Germany);
BMBF05P15WOFCA, DFG EClust 153, MLL, TU M\"unchen (Germany);
BMBF05P12RGGHM, JLU Giessen (Germany);
CNRS/IN2P3, IPN Orsay (France);
GACR13-06759S, MSMT LM2015049, Rez (Czech Republic);
BMBF05P15PXFCA, GSI WKAMPE1416, BU Wuppertal (Germany);
NCN 2013/10/M/ST2/00042 (Poland);
NSC 2016/23/P/ST2/04066 POLONEZ (Poland).

\section*{References}

\bibliography{mybibfile}

\end{document}